\documentstyle[sprocl,epsfig]{article}
\font\thinlinefont=cmr5
\input pictex.sty

\def\be{\begin{equation}}
\def\ee{\end{equation}}
\def\bea{\begin{eqnarray}}
\def\eea{\end{eqnarray}}

\newcommand{\bfi}[1]{\begin{figure}[#1]}
\newcommand{\efi}{\end{figure}}
\newcommand{\bpi}[2]{\begin{picture}(#1,#2)}
\newcommand{\epi}{\end{picture}}
\newcommand{\eref}[1]{(\ref{#1})}
\newcommand{\nn}{\nonumber\\}

\newcommand{\g}{\gamma}
\newcommand{\prop}{\Delta}
\newcommand{\propm}{{\Delta}_m}
\newcommand{\dsl}{\not \! \partial}
\renewcommand{\d}{\partial}
\newcommand{\T}{{\mathrm T}}
\newcommand{\B}{{\mathrm B}}
\newcommand{\A}{{\mathrm A}}

\def\OO{{\mathcal O}}

\begin{document}
\hspace*{8 cm}\parbox{4cm} {UG-FT-82/97\\
            November 1997}
                        \vspace*{2cm}

\title{Supersymmetric calculations with component fields \\ 
in differential renormalization
\footnote{Talk given at the International Workshop on Quantum
  Effects in MSSM, Universitat Aut\`onoma de Barcelona, September 1997} }

\author {F. del \'Aguila, $^{1}$
             A. Culatti, $^{1,2}$ 
             R. Mu\~noz Tapia, $^{1}$ and
             M. P\'erez-Victoria $^{1}$}
\address{$^{1}$Dpto. de F\'{\i}sica Te\'orica y del Cosmos,  
                       Universidad de Granada, \\ 
                                           18071 Granada, Spain; \\
                                           $^{2}$Dip. di Fisica, 
                       Universit\'a di Padova, 35131 Padova, Italy }  

\maketitle\abstracts{  
Two applications of the method of differential renormalization 
to superymmetric gauge theories are reviewed.
The photon propagator in supersymmetric QED is renormalized at 
one loop and the first supergravity contributions to the anomalous magnetic moment
of a charged lepton are obtained.}

\section{Introduction}
Quantum Field Theories are in general plagued with infinities,
which are usually regularized with some sort of cut-off. 
Then the renormalization program allows to get rid
of the singularities in a consistent way
by a redefinition of the parameters in the Lagrangian. In some
methods, like BPHZ, renormalization is  carried out
in one single step without intermediate regularization~\cite{bphz}.
Whatever scheme is used, it is desirable that it
maintains the symmetries of the theory under study. Theories with
a high degree of symmetry can be very demanding in this respect. 
This is the case of gauge and supersymmetric
theories. In a scalar theory with  very little symmetry, almost
any regulator does the job. 

Dimensional regularization~\cite{dimreg} is a
remarkable method that explicitly preserves gauge invariance.
Because of its relative simplicity and the importance
of gauge theories in the description of Nature, this method has 
become standard in many
computations in Quantum Field Theory. 
However, dimensional regularization has difficulties in coping with
supersymmetric theories: it can break supersymmetry (SUSY). 
Essentially, the reason  is that the equality of Bose and 
Fermi degrees of freedom 
only holds for specific values of the space-time dimension, which is changed
by this regularization.  
There exists a variant of the method, 
dimensional reduction ~\cite{dimred}, in which the field
components remain unchanged, while the loop integrals are performed
in a $d$-dimensional space.  This method preserves SUSY,
at least at lower orders, but at higher ones the situation is 
more involved, 
especially in the case of broken 
SUSY (see Ref.~\cite{drtj} and references therein).
Another approach
is based on higher derivative regularization supplemented by
Pauli-Villars~\cite{gaillard}. What is clear in any case
is the scarcity of
simple methods suitable for supersymmetric theories.

The method of differential renormalization (DR) has 
appeared recently~\cite{FJL}. It is a method of 
renormalization without regulators or explicit counterterms 
which works in coordinate space. 
DR does not modify the space-time dimension, thus being a 
candidate for preserving SUSY. 
The method has been shown to be quite powerful in a three-loop 
computation for the massless $\lambda \phi^4$ theory. 
Other applications include
lower dimensional theories, Chern-Simons, nonperturbative calculations, 
etc.~\cite{applications}. Formal aspects like checks of 
unitarity~\cite{counterterms}, its 
relation with dimensional regularization~\cite{Nuria}, 
the inclusion of masses~\cite{massive} 
and the consistency of the procedure to any 
order~\cite{systematic} have been also studied. 
Different versions of DR have been developed in Ref.~\cite{other}.
The first application to supersymmetric theories was  
a calculation of the $\beta $-function to three loops 
in the Wess-Zumino model~\cite{haagensen}. In Ref.~\cite{tesis} 
DR was applied to pure
supersymmetric gauge theories, and the $\beta $-function was 
obtained  to two loops and one loop, in the abelian and
non-abelian case, respectively. These calculations 
were performed with superspace techniques.
Very recently this method has also been employed 
in non-perturbative calculations in
supersymmetric gauge theories~\cite{AFGJ}. 
In the superspace formalism, SUSY is 
manifestly preserved; however the situation is more involved
in the physically interesting case of broken SUSY,
where it is usually preferred to work with component fields
(see, however, Ref.~\cite{spurion}). 
Here we review the two existing applications of
DR to supersymmetric gauge theories in the component
approach: the vacuum polarization in supersymmetric 
QED (SQED)~\cite{us} and 
the calculation of the anomalous 
magnetic moment of a charged lepton, $(g-2)_l$, in
unbroken supergravity~\cite{g2}. These two examples
provide non-trivial  tests of the potential of DR to
preserve (abelian) gauge invariance and SUSY.
In standard DR, the fulfilment of the corresponding
Ward identities is accomplished adjusting at the end
the different 
scales that appear in the renormalization procedure. 
In Ref.~\cite{us} we described a procedure 
to constrain the scales appearing in  DR 
in such a way that Ward 
identities are automatically satisfied. 
We have verified the consistency of this approach in different 
one-loop examples~\cite{us}. Here we shall use this
constrained procedure of DR.

In what follows we briefly review the method of DR and its constrained
form. Then we present the two one-loop supersymmetric calculations with 
component fields. The first example, 
the renormalization of the photon propagator in SQED, 
illustrates the method. Afterwards the more involved case of the evaluation 
of $(g-2)_l$ in unbroken supergravity is revised, and the results
are discussed. 
The last Section is devoted to conclusions.

\section{Differential renormalization and its constrained version}

In coordinate space the amplitudes are finite as long as point coordinates
are kept apart; singularities only arise when points coincide. These 
singularities, when too severe, give rise to divergences in 
integrals on internal points or in Fourier transforms.  In other words, 
the bare expressions are in general ill-defined distributions.
The idea of differential renormalization is to rewrite these expressions
as derivatives of less singular ones. The derivatives are understood
in the sense of distribution theory, i.e., acting formally by parts. 
The amplitudes written in this way 
are identical to the bare ones for separate points but behave well enough 
at coincident points. One has then the renormalized
amplitudes in coordinate space. In this process some dimensionful integration
constants appear that will play the role of renormalization scales. 
An illustrative example
is given in massless $\lambda \phi^4$ theory, where the four point
function at one loop has a factor $1/x^4$ ($x$ being the difference 
of any two external points). The renormalization is given by
the substitution
\be
\frac{1}{x^4} \rightarrow -\frac{1}{4}\Box\frac{\log(M^2 x^2)}{x^2}
\label{x4}.
\ee
The scale $M$ is the renormalization scale alluded to above.
Notice that the {\it r.h.s.}  in Eq.~\eref{x4} has a well defined
Fourier transform. We will work in Euclidean space, where the
handling of the expressions is simpler. 
Analogous equations can be obtained for other singularities
(see Ref.~\cite{FJL} for details) and the program can be in general
carried out for any theory at any number of loops~\cite{systematic}. 

In principle, the scales introduced for different diagrams are independent. 
These scales can be choosen in different ways (as long as the same
scale appears in identical subgraphs), each corresponding to a 
choice of the renormalization scheme. The easiest one is to 
take all scales to be equal. This choice is appropriate for a scalar  
$\lambda \phi^4$ theory, for instance. However, in theories with more 
symmetry much care must be taken when fixing the constants. In gauge
theories for example, 
there are  restrictions to be fulfilled which are dictated by 
the Ward identities.
The usual way of proceeding is to renormalize the amplitudes
and then relate the constants imposing the Ward identities.
Although this is a possible procedure, 
one would prefer that gauge symmetry were automatically preserved, as
occurs in dimensional regularization.

The constrained version of DR fixes the arbitrary constants from the 
beginning (see Ref.~\cite{us} for details). The idea is to restrict oneself 
to a set of consistent rules to manipulate the 
singular expressions. 
Besides the usual DR rules, differential equalities and formal
integration by parts, two other conditions appear to be
enough to conveniently fix the local terms at one loop. 
First, the factorization of delta-functions in the renormalization
procedure, i.e.,
\begin{equation}
\label{factorization}
[ F(x,x_1,\ldots,x_n)\delta(x-y)]^R=[ F(x,x_1,\ldots,x_n)]^R\delta(x-y),
\end{equation}
where $F$ is an arbitrary function and $R$ stands for ``renormalized''.
Second, one demands that the propagator equation
$(\Box -m^2) \Delta _m (x)=-\delta (x)$
has a general validity when 
embedded in any amplitude. This is to say that 
\begin{equation}
  \label{propagator}
  F(x,x_1,\ldots,x_n)(\Box-m^2)\Delta _m (x)=F(x,x_1,\ldots,x_n)[-\delta(x)]
\end{equation}
holds for $F$ arbitrary, and not only for well-behaved enough
functions. Above,  $\propm(x) = \frac{1}{4\pi ^2} \frac{m K_1(mx)}{x}$ 
is the Feynman propagator for
a particle of mass $m$, with $K_1$ a modified Bessel function.
These rules allow to renormalize a set of basic functions which 
are used to expand the amplitudes. The basic functions for tadpole, 
bubble and triangular diagrams with massless propagators are denoted by 
\begin{eqnarray}
  \A  & = & \prop(x) \delta(x)~, \label{A} \\
  \B[{\OO}] & = & \prop(x) {\OO}^x \prop(x) ~, \label{B}\\
  \T[\OO] & = & \prop(x) \prop(y) {\OO}^x \prop(x-y)~, \label{T} 
\end{eqnarray}
where $\prop(x)  =  \frac{1}{4\pi^2} \frac{1}{x^2}$ is
the massless propagator
and ${\OO}^x$ is a differential operator. 
For example, using these rules one can easily obtain 
\begin{eqnarray}
  \T^R[\Box] & = & - \B^R[1](x) \delta(x-y) \nn
            & = & \frac{1}{4} \frac{1}{(4\pi^2)^2} \Box
                  \frac{\log x^2 M^2}{x^2} \delta(x-y) ~.
\label{tbox}
\end{eqnarray}
The massive counterparts are defined accordingly and
will be denoted with a subindex $m$. The DR substitutions for
massive expressions can be obtained using recurrence
relations among Bessel functions (see Refs.~\cite{massive}
and~\cite{us} for a more detailed discussion). 
The only one we shall need is 
\begin{equation}
  \frac{m^2 K_1(m x)^2}{x^2} \rightarrow
  \frac{1}{2} (\Box - 4 m^2) \frac{m K_0(mx) K_1(mx)}{x}
  + \pi ^2 \log \frac{\bar{M}^2}{m^2} \delta(x)~ ,
\label{massidentity}
\end{equation}
where $\bar{M} = \frac{2M}{\gamma _E}$, and $\gamma _E = 1.781...$ 
is Euler's constant. 
When one has to handle tensor structures, the general procedure 
involves decomposing the expressions 
into trace and traceless parts. Consistency with
Eqs. (\ref{factorization}, \ref{propagator}) may imply the 
appearance of local terms. For instance, in the renormalization of 
$\T[\d_\mu \d_\nu]$ one has
\begin{equation}
\label{trace}
\T^R[\d_\mu\d_\nu] = \frac{1}{4} \delta_{\mu\nu} \T^R[\Box]
                    + \T[\d_\mu\d_\nu - \frac{1}{4} \delta_{\mu\nu}
                    \Box]
                   -  \frac{1}{128 \pi^2}  \delta(x) \delta(y)
                    \delta_{\mu\nu} ~.
\end{equation}
The last term is the local term alluded to, which is needed for the 
consistency of the equality:
\begin{eqnarray}
\B^R[\d_\mu](x) \delta(y)& = & - \d_\mu^x \Box^y \T[1]
  + \Box^y \T[\d_\mu] - 2 \d_\mu^x \d_\sigma^y \T[\d_\sigma]
  - \d_\mu^x \T^R[\Box] \nn
  && \mbox{}+ 2 \d_\sigma^y \T^R[\d_\mu\d_\sigma]
  + \T^R[\d_\mu \Box]~.
\label{relation}
\end{eqnarray}
To complete the renormalization of $\T[\d_\mu \d_\nu]$, one must note 
that only the 
first term in Eq.~\eref{trace} is singular and requires 
renormalization (see Eq.~\eref{tbox}). Eq.~\eref{trace} shows that in general
\begin{equation}
\delta_{\mu\nu} \T^{R}[\partial_\mu\partial_\nu]\neq
[\delta_{\mu\nu} \T[\partial_\mu\partial_\nu]]^{R}.
\label{notrace}
\end{equation}

\section{Vacuum polarization in SQED}

This is a simple example of a calculation in a supersymmetric
abelian gauge theory using component fields. 
As a matter of fact we will only check the transversality of the photon 
self-energy and not any SUSY relation.    
The vacuum polarization in SQED has two contributions,
one coming from the scalar loops (the corresponding diagrams are
depicted in Fig. 1), and the other from QED.
Before renormalizing, we will reduce 
their expressions to sums of basic functions and add the two parts.
The Feynman rules in coordinate space can be found in
Ref.~\cite{g2}.

\begin{figure}
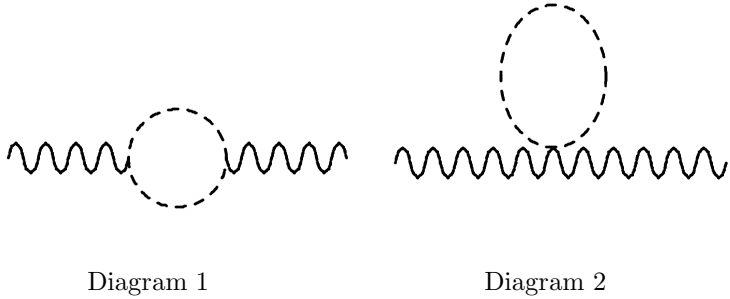

\begin{tabular}{cc}

\mbox{\beginpicture
\setcoordinatesystem units <1.04987cm,1.04987cm>
\unitlength=1.04987cm
\linethickness=0.5pt
\setplotsymbol ({\makebox(0,0)[l]{\tencirc\symbol{'160}}})
\setlinear
%
%
\setdashes < 0.1270cm>
\ellipticalarc axes ratio  0.620:0.620  360 degrees 
        from  3.715 24.765 center at  3.095 24.765
\setsolid
%
%
\plot  0.953 24.765      0.975 24.855
         0.997 24.914
         1.048 24.955
         1.107 24.896
         1.126 24.835
         1.143 24.765
         1.160 24.695
         1.179 24.634
         1.238 24.575
         1.297 24.634
         1.317 24.695
         1.334 24.765
         1.350 24.835
         1.370 24.896
         1.429 24.955
         1.488 24.896
         1.507 24.835
         1.524 24.765
         1.541 24.695
         1.560 24.634
         1.619 24.575
         1.678 24.634
         1.698 24.695
         1.715 24.765
         1.731 24.835
         1.751 24.896
         1.810 24.955
         1.869 24.896
         1.888 24.835
         1.905 24.765
         1.922 24.695
         1.941 24.634
         2.000 24.575
         2.059 24.634
         2.079 24.695
         2.095 24.765
         2.112 24.835
         2.132 24.896
         2.191 24.955
         2.250 24.896
         2.269 24.835
         2.286 24.765
         2.303 24.695
         2.322 24.634
         2.381 24.575
         2.432 24.616
         2.454 24.675
         2.477 24.765
        /
%
%
\plot  3.715 24.765      3.739 24.676
         3.761 24.617
         3.810 24.575
         3.870 24.634
         3.889 24.695
         3.906 24.765
         3.923 24.835
         3.942 24.896
         4.000 24.955
         4.060 24.895
         4.080 24.835
         4.097 24.765
         4.113 24.695
         4.132 24.635
         4.191 24.575
         4.251 24.634
         4.270 24.695
         4.287 24.765
         4.304 24.835
         4.323 24.896
         4.381 24.955
         4.441 24.895
         4.461 24.835
         4.478 24.765
         4.494 24.695
         4.513 24.635
         4.572 24.575
         4.632 24.634
         4.651 24.695
         4.668 24.765
         4.685 24.835
         4.704 24.896
         4.763 24.955
         4.822 24.895
         4.842 24.835
         4.859 24.765
         4.875 24.695
         4.894 24.635
         4.953 24.575
         5.013 24.634
         5.032 24.695
         5.049 24.765
         5.066 24.835
         5.085 24.896
         5.143 24.955
         5.194 24.913
         5.216 24.855
         5.239 24.765
        /
\linethickness=0pt
\putrectangle corners at  0.0 26.0 and  5.0 23.5
\endpicture}&
\mbox{\beginpicture
\setcoordinatesystem units <1.04987cm,1.04987cm>
\unitlength=1.04987cm
\linethickness=0.5pt
\setplotsymbol ({\makebox(0,0)[l]{\tencirc\symbol{'160}}})
\setshadesymbol ({\thinlinefont .})
\setlinear
%
%
\setdashes < 0.1270cm>
\ellipticalarc axes ratio  0.667:0.904  360 degrees 
        from  3.619 25.576 center at  2.953 25.576
\setsolid
%
%
\plot  2.477 24.479      2.499 24.569
         2.521 24.628
         2.572 24.670
         2.631 24.610
         2.650 24.549
         2.667 24.479
         2.684 24.409
         2.703 24.348
         2.762 24.289
         2.821 24.348
         2.841 24.409
         2.857 24.479
         2.874 24.549
         2.894 24.610
         2.953 24.670
         3.012 24.610
         3.031 24.549
         3.048 24.479
         3.065 24.409
         3.084 24.348
         3.143 24.289
         3.202 24.348
         3.222 24.409
         3.239 24.479
         3.255 24.549
         3.275 24.610
         3.334 24.670
         3.393 24.610
         3.412 24.549
         3.429 24.479
         3.446 24.409
         3.465 24.348
         3.524 24.289
         3.583 24.348
         3.603 24.409
         3.620 24.479
         3.636 24.549
         3.656 24.610
         3.715 24.670
         3.774 24.610
         3.793 24.549
         3.810 24.479
         3.827 24.409
         3.846 24.348
         3.905 24.289
         3.956 24.330
         3.978 24.389
         4.000 24.479
        /
%
%
\plot  4.000 24.479      4.023 24.569
         4.045 24.628
         4.096 24.670
         4.155 24.610
         4.174 24.549
         4.191 24.479
         4.208 24.409
         4.227 24.348
         4.286 24.289
         4.345 24.348
         4.365 24.409
         4.381 24.479
         4.398 24.549
         4.418 24.610
         4.477 24.670
         4.536 24.610
         4.555 24.549
         4.572 24.479
         4.589 24.409
         4.608 24.348
         4.667 24.289
         4.726 24.348
         4.746 24.409
         4.763 24.479
         4.779 24.549
         4.799 24.610
         4.858 24.670
         4.917 24.610
         4.936 24.549
         4.953 24.479
         4.970 24.409
         4.989 24.348
         5.048 24.289
         5.099 24.330
         5.121 24.389
         5.143 24.479
        /
%
%
\plot  0.953 24.479      0.975 24.569
         0.997 24.628
         1.048 24.670
         1.107 24.610
         1.126 24.549
         1.143 24.479
         1.160 24.409
         1.179 24.348
         1.238 24.289
         1.297 24.348
         1.317 24.409
         1.334 24.479
         1.350 24.549
         1.370 24.610
         1.429 24.670
         1.488 24.610
         1.507 24.549
         1.524 24.479
         1.541 24.409
         1.560 24.348
         1.619 24.289
         1.678 24.348
         1.698 24.409
         1.715 24.479
         1.731 24.549
         1.751 24.610
         1.810 24.670
         1.869 24.610
         1.888 24.549
         1.905 24.479
         1.922 24.409
         1.941 24.348
         2.000 24.289
         2.059 24.348
         2.079 24.409
         2.095 24.479
         2.112 24.549
         2.132 24.610
         2.191 24.670
         2.250 24.610
         2.269 24.549
         2.286 24.479
         2.303 24.409
         2.322 24.348
         2.381 24.289
         2.432 24.330
         2.454 24.389
         2.477 24.479
        /
\linethickness=0pt
\putrectangle corners at  0.5 27.3 and 5.2 23.272
\endpicture} \\
\ \ \ \ Diagram 1 & Diagram 2
\end{tabular}
\caption{One-loop diagrams contributing to the 
vacuum polarization in scalar QED. \label{fig}}
\end{figure}

The vacuum polarization in scalar QED reads
\begin{eqnarray}
  \Pi_{\mu\nu}^{(1)}(x) & = & 
  - e^2 \propm(x) 
  \stackrel{\leftrightarrow}{\d}_\mu
  \stackrel{\leftrightarrow}{\d}_\nu 
  \propm(x) ~, \\
  \Pi_{\mu\nu}^{(2)}(x) & = &  
  -2 e^2 \delta_{\mu\nu} \propm(x) \delta(x) ~,
\end{eqnarray}
where
$\A \stackrel{\leftrightarrow}{\d} \B= \A \d \B - \B \d \A$. 
In terms of basic functions one has
\begin{eqnarray}
  \Pi_{\mu\nu}^{(1)}(x) & = & -e^2 \{ 4\B_m[\d_\mu\d_\nu]
    - \d_\mu\d_\nu B_m[1] \} ~, \\
  \Pi_{\mu\nu}^{(2)}(x) & = & -2 e^2 \delta_{\mu\nu} \A_m ~.
\label{tadpole}
\end{eqnarray}
Now the propagator equation can be used to rewrite the tadpole
contribution, Eq. \eref{tadpole}, in terms of bubble functions, 
\begin{eqnarray}
  \A_m & = & \propm(x) \delta(x) \nn 
     & = & -\propm(x) (\Box-m^2) \propm(x) \nn  
     & = & - \B_m[\Box]+m^2 \B_m[1] ~.
\end{eqnarray}
On the other hand the QED contribution (diagram 1 in Fig. 1, but with 
a fermion running in the loop) can be written 
\begin{equation}
  \Pi_{\mu\nu}(x) = 4 e^2 \{ (m^2 \delta_{\mu\nu}
  + \frac{1}{2} \delta_{\mu\nu} \Box - \d_\mu\d_\nu) \,
  \B_m[1] + 2 \B_m[\d_\mu\d_\nu] - \delta_{\mu\nu} 
  \B_m[\Box] \}. 
\end{equation} 
Then the supersymmetric QED  vacuum polarization is the
sum of the spinor diagram plus twice  
the scalar ones. One directly obtains 
a transverse result depending only on 
one basic function: 
\begin{eqnarray}
  \Pi_{\mu\nu}(x) = -2 e^2 (\d_\mu\d_\nu - \delta_{\mu\nu} \Box)
  \B_m[1]~.
\end{eqnarray}
This equation exhibits the consistency of 
constrained DR and abelian gauge invariance in this simple 
one-loop supersymmetric calculation. 
The renormalization is completed substituting $\B_m[1]$ by 
its renormalized expression, given by Eq.~(\ref{massidentity}): 
\begin{eqnarray}
  \Pi_{\mu\nu}^R(x) & = & - \frac{ e^2}{(4\pi^2)^2} 
  (\d_\mu\d_\nu - \delta_{\mu\nu} \Box) [
  (\Box - 4m^2) \frac{m K_0(mx) K_1(mx)}{x} \nn
  && \mbox{} + 2 \pi^2 \log \frac{\bar{M}^2}{m^2} 
  \delta(x) ]  ~.
\end{eqnarray}

\section{Supergravity contributions to $(g-2)_l$}

This calculation is more involved. 
In order to show how coordinate space techniques can be used
to calculate a quantity typically defined for fixed momenta, we 
first  work out the standard 
QED correction in coordinate space.

\subsection{$(g-2)_l$ in momentum and coordinate space}

The anomalous magnetic moment of the electron,
$(g-2)_l$, is usually defined
in momentum space as a static limit.
For $p$ and $p'$ being the incoming momenta
of the electrons and  $q=p+p'$ being the outgoing momentum of
the photon,  the parity conserving vertex
containing all radiative corrections 
can be expressed, for on-shell external electrons,
in terms of two form factors: 
\begin{equation}
  \Lambda_\mu(p,p') =  i e [(F_1(q^2) + F_2(q^2)) \gamma_\mu
  + \frac{F_2(q^2)}{2m} (p_\mu - p_\mu') ] ~.
\label{formfactors}
\end{equation}
Then the anomalous magnetic moment is defined as
\be
  \frac{g-2}{2} = \lim_{q^2 \rightarrow 0} F_2(q^2) .
\ee
The corresponding expression for the vertex in coordinate space 
is related to the momentum space one via
\begin{equation}
  \Lambda_\mu(p,p') = 
  \int d^4 x  d^4 y e^{ip\cdot x} e^{i p' \cdot y}
  \Lambda_\mu(x,y) ~.
\label{fourier}
\end{equation}
The on-shell condition for the external electrons corresponds
to imposing Dirac equation on their wave functions. This means
that we must substitute terms of the form 
$\dsl^x f(x,y)$ and
$f(x,y) \stackrel{\leftarrow}{\dsl^y}$ in 
$\Lambda_\mu(x,y)$ by $-m f(x,y)$ and $m f(x,y)$, 
respectively.
Similarly, the static limit corresponds in coordinate
space to imposing $\Box A_\mu = 0$. Therefore, terms in
$\Lambda_\mu(x,y)$ like
$\Box^z f(x,y) = [-(\d_\mu^x + \d_\mu^y)]
[-(\d_\mu^x + \d_\mu^y)] f(x,y)$  
must be neglected. 
In practice, however, one cannot
always extract all derivatives, and pieces containing
internal derivatives remain. Hence, Dirac and
Maxwell equations cannot be directly imposed. The
simplest solution is to perform at the end a Fourier transform 
to obtain $\Lambda_\mu(p,p')$ and to take the appropriate 
limits. This procedure may seem to end up in
the usual momentum space one. The difference, however, is that 
renormalization is
carried out before Fourier transforming, so the Fourier
integrals  are finite and can be computed in four dimensions without 
any regulator. 

In QED, $(g-2)_l$ vanishes at tree level (this is predicted by
Dirac's equation: $g=2$ for a particle of spin $1/2$), but it is well-known 
that a finite non-zero
$(g-2)_l$ is generated at one loop. At this order the vertex correction,
 $V_\mu$, 
is given by the standard triangular diagram and reads 
\begin{equation}
   V_\mu(x,y) = (-ie)^3 \g_\alpha (\dsl^x
   - m) \propm(x) \g_\mu
   (- \dsl^y - m) \propm(y)
   \g_\alpha \prop(x-y) .
\label{v1}
\end{equation}
Using Leibnitz rule to rearrange
derivatives, 
Eq.~\eref{v1} can be expressed in terms of the  
triangular functions defined in Eq.~\eref{T}, in this case with 
two massive propagators, i.e.,
\begin{equation}
  \T_m[ {\OO} ] = \propm(x) \propm(y) {\OO}^x \prop(x-y)~.
\end{equation}
Using Dirac's equation whenever possible the resulting 
expression is
\begin{eqnarray}
  V_\mu(x,y) =  ie^3 && \{ -4 \g_\mu \d^x \cdot \d^y \T_m[1]
  + 4 (\d_\mu^x - \d_\mu^y) \T_m[\dsl] - 4 \g_\mu (\d_a^x-
  \d_a^y) \T_m[\d_a]  \nonumber \\
  && \mbox{} + 4 m \T_m[\d_\mu] + 2 \g_\mu \T_m[\Box]   
  - 4 \g_a \T_m[\d_a\d_\mu] \} .
\end{eqnarray}
The terms proportional to $\g_\mu$ do not contribute
to $(g-2)_l$ and we ignore them in the following. Of the rest,
only the last triangular function, $\T_m[\d_a\d_\mu]$,
is singular. As in Eq.~\eref{trace} it is
decomposed into a part proportional to $\delta_{\mu\nu}$
and a finite traceless part. Only this traceless part 
contributes to $(g-2)_l$. Hence, $(g-2)_l$ can
be extracted from the finite expression,
\begin{equation}
  V_\mu(x,y)^{g-2} = ie^3 \{-4 (\d_\mu^y - \d_\mu^x) \T_m[\dsl]
  + 4m \T_m[\d_\mu] - 4 \g_a 
  \T_m[\d_a\d_\mu - \frac{1}{4} \delta_{a\mu} \Box] \}~,
\end{equation}
which can be readily Fourier transformed (see Eq.~\eref{fourier}).
The necessary integrals in the static limit,
\begin{equation}
  p^2= p'^2 = -m^2 ~~, ~~q^2 \rightarrow 0 ~,
\end{equation}
are
\begin{eqnarray}
  \hat{\T}_m[\d_\mu] & = & - \frac{i}{32\pi^2 m^2}
    (p_\mu - p_\mu') , \\
  \hat{\T}_m[\d_a\d_\mu - \frac{1}{4} \delta_{a\mu} \Box]
    & = & - \frac{i}{32\pi^2 m^2} \{ -\frac{1}{6}
    (p_a p_\mu' + p_a' p_\mu) + 
    \frac{1}{3} (p_a p_\mu + p_a' p_\mu') \nonumber \\
    && \hspace{2cm} \mbox + \frac{1}{4} m^2 \delta_{a\mu} \}~,
\end{eqnarray}
Now the value of the anomalous magnetic
moment can be read from the coefficient of $p_\mu-p_\mu'$ in
$V_\mu(x,y)^{g-2}$
(see Eq.~\eref{formfactors}):
\begin{equation}
  \frac{g-2}{2} = \frac{2m}{ie} \, 
  \frac{ie^3}{16\pi^2 m}(4-4\frac{1}{2}-4\frac{1}{4}) = \frac{\alpha}{2\pi}~.
\end{equation}
This is the well-known Schwinger result~\cite{schwinger}.

\subsection{$(g-2)_l$ in unbroken supergravity}

In a supersymmetric theory $(g-2)_l$ 
vanishes because no such term appears in the Lagrangian of a chiral 
supermultiplet ~\cite{ferrara}. 
Hence, as long as SUSY is preserved, all quantum
corrections must cancel order by order. 
Therefore, the anomalous magnetic moment of the
lepton besides being an observable, is also an ideal arena 
to check theoretical implications and to perform consistency tests 
of regularization methods. 
Ferrara and Remiddi also proved explicitly  that in global SUSY the 
one--loop QED corrections, order $e^3$, do cancel.
This is to say that the fermion contribution (Schwinger result) and the 
scalar one cancel each other. The latter results from twice the same 
triangular diagram but with the slepton and the photino 
replacing the lepton and the photon, respectively.

The one--loop gravitational
corrections are of order $e\kappa^2= 8\pi e G_N$, resulting from
a graviton or gravitino exchange~\cite{sugra1} (the corresponding 
diagrams are depicted in Fig. 2). 
\begin{figure}
\centering
\mbox{%
\epsfig{file=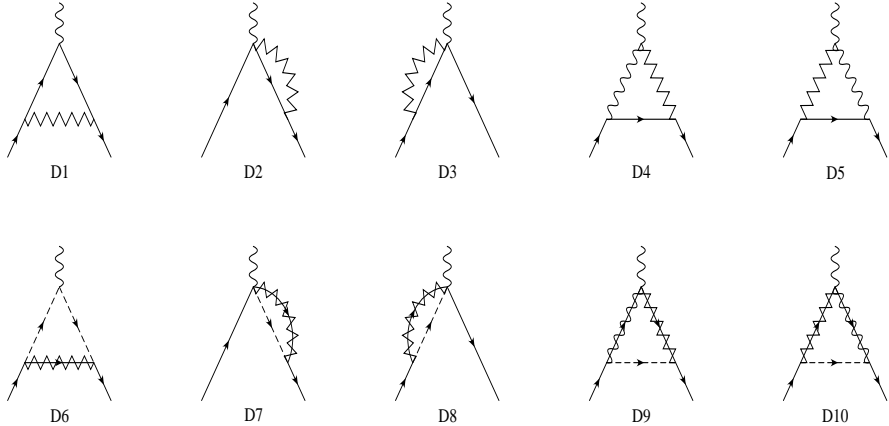 
        ,height=6cm  
        ,width=11.7cm   
       }%
 }
\caption{Diagrams of order $e \kappa^2 $ contributing to $(g-2)_l$ 
in supergravity. 
A graviton is exchanged in diagrams D1-D5 and a gravitino
in D6-D10.} 
\label{fig-sugra}      
\end{figure}
Using dimensional
regularization,
Berends and Gastmans~\cite{berends} calculated  
the five diagrams where a graviton
is exchanged. All five diagrams are infinite but
their sum is finite. The finiteness of $(g-2)_l$ 
in a non--renormalizable theory such as gravity
seemed miraculous. Del Aguila et al.~\cite{paco} and 
Bellucci et al.~\cite{bellucci}
checked that when gravitation is embedded in a supersym\-metric theory
(unbroken), the
contributions from the graviton and the gravitino cancel, as 
required by SUSY. 
Bellucci et al.~\cite{bellucci} also 
traced back to an effective chiral symmetry in the 
gravitino sector the finiteness of the gravitino contribution 
and then of the graviton contribution, if their sum has to vanish. 
Dimensional regularization does not yield a vanishing 
value for $(g-2)_l$. This is one example of a case where 
dimensional regularization
breaks SUSY. 
A (one--loop) SUSY preserving method  is  required 
in order to obtain 
such a cancellation and this 
was shown to be the case in dimensional reduction. 
In Ref.~\cite{g2} we calculated these contributions using DR
(the 
Lagrangian, Feynman rules and other technical details can be found there).
The use of the rule extending the validity of 
the propagator equation allowed us to relate diagrams with
different topology before explicit renormalization. 
Then considering the sum of each diagram and its supersymmetric partner, 
we found that the contibution of each sum to $(g-2)_l$ depends only 
on one singular scalar basic function, $\T_m[\Box]$, apart from
finite terms. (This is analogous to what happens in the calculation  
of the vacuum polarization in SQED above.) 
The terms proportional to $\T_m[\Box]$ cancel out in the complete sum 
and so do the finite terms. 
In this way SUSY is preserved, i.e., a vanishing value of $(g-2)_l$ is 
obtained. Terms proportional to $q_\mu$ did not appear, either. Hence, 
on-shell gauge invariance, forbidding these terms, is also respected 
in this calculation. 
It is worth to note, however, that singular basic functions with other 
tensor structure do appear in individual graphs and in the total graviton 
(gravitino) contribution. Then, for these contributions which  
become physically meaningful if SUSY is broken, the tensor 
decomposition of these singular functions and the local terms they 
introduce (see for instance Eq. \eref{trace}) are relevant. In Ref.~\cite{g2}
we used the engineering trace-traceless decomposition neglecting local 
terms. We call this method `partially constrained' DR. 
Therefore, the partial results may and do differ when using 
`partially constrained' DR or constrained DR (for the latter includes 
the local terms), although the total sum is the same. 
In Table \ref{tab:g-2} we gather the results in
dimensional regularization, dimensional reduction, 
`partially constrained' DR and constrained DR.
As can be observed, the last three methods (columns) preserve SUSY 
(add to zero), but `partially constrained' DR
gives a different value for the graviton (gravitino) 
contribution. 
The graviton (gravitino) contribution depends on the regularization 
method and is not well defined by itself. This is related to the
fact that we are dealing with a non-renormalizable theory, and in
the non-supersymmetric case no symmetry protects the value of $(g-2)_l$.
The fact that dimensional reduction and constrained DR give the
same result for this contribution seems to indicate that the requirement that
the regularization/renormalization method be compatible with
gauge invariance and SUSY greatly constrains
the results at one loop.
\newpage
\begin{table}[h]
\caption{Contributions of the diagrams in Fig.~2  to 
$\left( \frac{g-2}{2}\right)_l$ in units of 
$\frac{G_N m^2}{\pi}$, obtained with dimensional regularization, 
dimensional reduction, `partially constrained' DR and constrained DR.
\label{tab:g-2}}


\begin{center}

\footnotesize

\begin{tabular}{|l| l |l |l |l|}
\hline  
& \multicolumn{1}{|c|}{Dimensional} & \multicolumn{1}{|c|}{Dimensional} &
\multicolumn{1}{|c|}{`Partially} & \\
\multicolumn{1}{|c|}{Diagram} & \multicolumn{1}{|c|}{Regularization} &
\multicolumn{1}{|c|}{Reduction} & \multicolumn{1}{|c|}{Constrained' DR} &
\multicolumn{1}{|c|}{Constrained  DR} \\  
        \hline  & & & & \\
D1      & $\frac{1}{3} \frac{1}{n-4}-\frac{61}{36}$
        & $\frac{1}{3}\frac{1}{n-4}-\frac{29}{18}$ 
        & $-\frac{1}{6} \log\left(\frac{\bar{M}^2}{m^2}\right) -\frac{25}{18}$ 
        & $-\frac{1}{6} \log\left(\frac{\bar{M}^2}{m^2}\right)
        -\frac{25}{18}$\\ 
        & & & & \\
D2+D3   & $\frac{11}{3} \frac{1}{n-4}-\frac{32}{9}$
        & $\frac{11}{3} \frac{1}{n-4}-\frac{35}{9}$ 
        & $-\frac{11}{6} \log \left(\frac{\bar{M}^2}{m^2}\right)-\frac{11}{18}$
        & $-\frac{11}{6} \log
        \left(\frac{\bar{M}^2}{m^2}\right)-\frac{1}{9}$\\
        & & & & \\
D4+D5   & $-4\frac{1}{n-4}+7$
        & $-4\frac{1}{n-4}+6$ 
        & $ 2  \log \left(\frac{\bar{M}^2}{m^2}\right) +1$
        & $ 2  \log \left(\frac{\bar{M}^2}{m^2}\right) +2$\\ 
        & & & &    \\
Graviton  & & & & \\     
(D1+D2+D3 &$7/4$              &1/2          &$-1$   &1/2\\
+D4+D5) & & & & \\ 
        & & & &\\ \hline & & & & \\
D6      &$\frac{8}{3}  \frac{1}{n-4}-\frac{55}{18}$
        &$\frac{8}{3}  \frac{1}{n-4}-\frac{37}{18}$ 
        &$-\frac{4}{3} \log \left(\frac{\bar{M}^2}{m^2}\right) +\frac{19}{18}$
        &$-\frac{4}{3} \log\left(\frac{\bar{M}^2}{m^2}\right)+\frac{19}{18}$\\ 
        & & & &   \\
D7+D8      &$\frac{4}{3}  \frac{1}{n-4}-\frac{13}{9}$
           &$\frac{4}{3}  \frac{1}{n-4}-\frac{4}{9}$ 
           &$-\frac{2}{3} \log \left(\frac{\bar{M}^2}{m^2}\right)+\frac{17}{18}$
           &$-\frac{2}{3} \log\left(\frac{\bar{M}^2}{m^2}\right)+\frac{4}{9}$\\
        & & & & \\
D9+D10     &$-4\frac{1}{n-4}+4$
           &$-4\frac{1}{n-4}+2$ 
           &$ 2  \log \left(\frac{\bar{M}^2}{m^2}\right) -1$
           &$ 2  \log \left(\frac{\bar{M}^2}{m^2}\right) -2$\\
        & & & & \\
Gravitino   &  & & &   \\ 
(D6+D7+D8  &$-1/2$             &$-1/2$          &$1$  &$-1/2$\\
+D9+D10) &  & & &   \\ 
        & & & &\\  \hline 
TOTAL    &   &  &  & \\
(Graviton & 5/4 &0 &0 &0\\ +Gravitino) & & & &  \\
 \hline
\end{tabular}
\end{center}
\end{table}

\section{Conclusions}

DR is a method of renormalization recently proposed~\cite{FJL}, 
which works in coordinate space and does not introduce any 
intermediate regulator.
It seems to have the potential of 
preserving gauge and chiral invariance. 
This procedure has been applied in several contexts 
and in particular to several supersymmetric calculations 
with component fields, which we have reviewed here. The method 
can be constrained to get rid of arbitrary constants, except 
for the neccesary renormalization group scale. 
Constrained DR is defined by a set of rules which completely fix 
the renormalization of singular expressions at least to one loop. 
In the two examples worked out, a transverse renormalized 
vacuum polarization in  SQED and a vanishing $(g-2)_l$ in supergravity
have been obtained. The use of the rule extending 
the validity of the propagator equation plays an essential role: 
it allows to relate the expressions appearing in different graphs 
and enforce the 
supersymmetric and gauge invariance constraints from the beginning. 
In both cases, SUSY cancellations make the result 
insensitive to the renormalization of basic
functions with non-trivial tensor structure, and thus 
to the inclusion of local terms in the tensor decomposition.
From only these two calculations, however, it cannot be said how general this
effect is. At any rate, 
if SUSY is broken it is clear that the local terms
become relevant. As shown in Ref. ~\cite{us}, the extensive
use of the rules completely determines these terms, so 
constrained DR can in principle be used in the case of broken 
SUSY.
Here we have only presented a simple (but non-trivial)  
consistency check of the constrained DR method for 
supersymmetric abelian gauge theories. 
A real test (or proof) should
consider the Ward identities of both SUSY and
gauge invariance, as well as higher orders.

\section*{Acknowledgements}

This work has been supported by CICYT, contract
number AEN96-1672, and by Junta de Andaluc\'{\i}a, 
FQM101.
RMT and MPV thank MEC for financial support.
FA thanks the Universitat Aut\`onoma de Barcelona
for its hospitality.



\section*{References}
\begin{thebibliography}{99}

\bibitem{bphz} N.N. Bogoliubov and O.Parasiuk, Acta Math. {\bf 97} 
             (1957) 227; 
             K. Hepp, Comm. Math. Phys. {\bf 2} (1966) 301;
             W. Zimmermann,  Comm. Math. Phys. {\bf 15} (1969) 208;
             E. Corrigan, P. Goddard, H. Osborn and S Templeton,
             Nucl. Phys. {\bf B159} (1979) 469.
\bibitem{dimreg} G. 't  Hooft and M. Veltman, Nucl. Phys. {\bf B44} 
                 (1972) 189; 
                 C.G. Bollini and J. Giambiagi, Nuovo Cim. {\bf 12 B}
                 (1972) 20; J.F. Ashmore, Nuovo Cim. Lett. {\bf 4}
                 (1972) 289; 
                 G.M. Cicuta and E. Montaldi, Nuovo Cim. Lett. {\bf 4}
                 (1972) 329.
\bibitem{dimred} W. Siegel, Phys. Lett. {\bf B84} (1979) 193;
                 Phys. Lett. {\bf B94} (1980) 37.
\bibitem{drtj} I. Jack and D.R.T. Jones, Liverpool University preprint
                LTH 400, hep-ph/9707278, to appear in `Perspectives
				 on Supersymmetry', World Scientific, 
                                 Ed. G. Kane.
\bibitem{gaillard} P. West, Nucl. Phys. {\bf B268} (1986) 113;
                   M.K. Gaillard, Phys. Lett. {\bf B342}, 125 (1995);
                   Phys. Lett. {\bf B347}, 284 (1995).
\bibitem{FJL} D.Z. Freedman, K. Johnson and J.I. Latorre, Nucl. Phys.
              {\bf B371}  (1992) 353.
\bibitem{applications} R. Mu\~noz-Tapia, Phys. Lett. {\bf B295} (1992) 95;
                  D.Z. Freedman, G. Grignani, K. Johnson and N. Rius, 
                  Ann. Phys. {\bf 218} (1992) 75;
                  P.E. Haagensen and J.I. Latorre, 
                  Ann. Phys. (N.Y.) {\bf 221} (1993) 77;
                  C. Manuel, Int. J. Mod. Phys. {\bf A8} (1993) 3223;
                  D.Z. Freedman, G. Lozano and N. Rius, Phys. Rev. {\bf D49}
                  (1994) 1054;
                  J. Comellas, P.E. Haagensen and J.I. Latorre, 
                  Int. J. Mod. Phys. {\bf A10} (1995) 2819; 
				  M. Chaichian, W.F. Chen and H.C. Lee, 
                  hep-th/9703219 v2, to appear in
                  Phys. Lett. {\bf B}.
\bibitem{Nuria}  G. Dunne and N. Rius, Phys. Lett. {\bf B293} (1992) 367.
\bibitem{massive} P.E. Haagensen and J.I. Latorre, Phys. Lett. 
                  {\bf B283} (1992) 293.
\bibitem{counterterms} D.Z. Freedman, K. Johnson, R. Mu\~noz-Tapia and X. 
                       Vilasis-Cardona, Nucl. Phys. {\bf B395} (1993) 454.
\bibitem{systematic} J.I. Latorre, C. Manuel and X. Vilasis-Cardona,
                     Ann. Phys. {\bf 231} (1994) 149.
\bibitem{other} V.A. Smirnov, Theor. Math. Phys. {\bf 96} (1993) 974;
                Nucl. Phys. {\bf B427} (1994) 325;
                Z. Phys. {\bf C67} (1995) 531;
                Theor. Math. Phys. {\bf 108} (1997) 953;
                O. Schnetz, J. Math. Phys. {\bf 38} (1997) 738. 
\bibitem{haagensen} P.E. Haagensen, Mod.  Phys. Lett. {\bf A7} (1992) 893.
\bibitem{tesis} Yun S. Song, Ph.D. thesis.
\bibitem{AFGJ}	D. Anselmi, D.Z. Freedman, M.T. Grisaru,
                  A.A. Johansen, BRX-TH-420, CPTH-S.553.0897,
                  HUTP-97/A037, MIT-CTP-2666, hep-th/9708042.
\bibitem{spurion} L. Girardello and M.T. Grisaru, Nucl. Phys.
                  {\bf B194} (1982) 65.
\bibitem{us} F. del Aguila, A. Culatti, R. Mu\~noz Tapia and
                M. P\'erez-Victoria, UG-FT-73/97, KA-TP-10-1997, 
                DFPD 97/TH/38,
                hep-th/9709067, to appear in Phys. Lett.  {\bf B};
                F. del Aguila and M. P\'{e}rez-Victoria, UG-FT-81/97,
                hep-ph/9710442, to appear in Acta Physica Polonica {\bf B}.
\bibitem{g2} F. del Aguila, A. Culatti, R. Mu\~noz Tapia and
                M. P\'erez-Victoria, Nucl. Phys. {\bf B504}
                (1997) 532.
\bibitem{schwinger} J. Schwinger, Phys. Rev. {\bf 73} (1948) 416.                    
                 
\bibitem{ferrara}  S. Ferrara and E. Remiddi, Phys. Lett. {\bf B53} (1974) 347;
\bibitem{sugra1}   P. van Nieuwenhuizen, Phys. Rep. {\bf C68} (1981) 189.
\bibitem{berends}  F.A. Berends and R. Gastmans,  Phys. Lett. {\bf B55}
                   (1975) 311.
\bibitem{paco}     F. del Aguila, A. Mendez and F.X. Orteu, Phys. Lett.
                    {\bf B145} (1984) 70.
\bibitem{bellucci}  S. Bellucci, H. Cheng and S. Deser, Nucl. Phys. 
                    {\bf B252} (1985) 389. 
\end{thebibliography}
\end{document}